\documentstyle[11pt,aaspp4,epsfig]{article}
\slugcomment{Accepted by The Astrophysical Journal Letters on 5/19/98}

\begin{document}

\title{Evolution of AQL X-1 During the Rising Phase of its 1998 Outburst}
\author{Wei Cui\altaffilmark{1}, Didier Barret\altaffilmark{2}, S.~N.~Zhang\altaffilmark{3,4}, Wan Chen\altaffilmark{5,6}, Laurence Boirin\altaffilmark{2}, and Jean Swank\altaffilmark{5}}

\altaffiltext{1}{Room 37-571, Center for Space Research, Massachusetts Institute of Technology, Cambridge, MA 02139}

\altaffiltext{2}{Centre d'Etude Spatiale des Rayonnements, 9 Av du Colonel Roche, 31028 Toulouse Cedex 04, FRANCE}

\altaffiltext{3}{ES-84, NASA/Marshall Space Flight Center, Huntsville, AL 35812}

\altaffiltext{4}{also Universities Space Research Association}

\altaffiltext{5}{NASA/Goddard Space Flight Center, Code 661, Greenbelt, MD 20771}

\altaffiltext{6}{also Department of Astronomy, University of Maryland, College Park, MD 20742}

\begin{abstract}
We present results from 16 snapshots of Aql X-1 with the {\it Rossi X-ray 
Timing Explorer}during the rising phase of its recent outburst. The 
observations
were carried out at a typical rate of once or twice per day. The source shows
interesting spectral evolution during this period. Phenomenologically, it 
bears remarkable similarities to ``atoll'' sources. Shortly after the onset 
of the outburst, the source is seen to be in an ``island'' state, but with 
little X-ray variability. It then appears to have made a rapid spectral 
transition (on a time scale less than half a day) to another ``island'' state,
where it evolves slightly and stays for 4 days. In this state, the observed
X-ray flux becomes increasingly variable as the source brightens. Quasi-period
oscillation (QPO) in the X-ray intensity is detected in the frequency range 
670--870 Hz. The QPO frequency increases with the X-ray flux while 
its fractional rms decreases. The QPO becomes undetectable following
a transition to a ``banana'' state, where the source continues its evolution 
by moving up and down the ``banana'' branch in the color-color diagram as the 
flux (presumably, the mass accretion rate) fluctuates around the peak of the 
outburst. Throughout the entire period, the power density spectrum is 
dominated by very-low frequency noises. Little power can be seen above 
$\sim$1 Hz, which is different from typical ``atoll'' sources. In the 
``banana'' state, the overall X-ray variability remains low (with fractional
rms $\sim$3--4\%) but roughly constant.
The observed X-ray spectrum is soft with few photons from above $\sim$25 keV,
implying the thermal origin of the emission. The evolution of both 
spectral and temporal X-ray properties is discussed in the context of 
disk-instability models.
\end{abstract}


\section{Introduction}
Soft X-ray transients (SXTs) constitute an important subclass of low-mass 
X-ray binaries. For most of the time, they appear as extremely faint X-ray
sources, if detected at all, but occasionally they brighten up by orders 
of magnitude in X-ray intensity, becoming the brightest X-ray sources in
the sky in some cases. There appears to be some consensus that thermal 
instability causes a sudden surge in the mass accretion rate through the 
disk and, thus, initiates an X-ray outburst (review by King 1995 and 
references therein). Archival databases have been established over the years
for the study of SXTs in outbursts and have been continually enriched by the
on-going missions. Few observations, however, have been made of SXTs during
the rising phase of an outburst because of the difficulty in catching such a
brief period (typically lasting for a few days; cf., Chen, 
Shrader, \& Livio 1997). The situation has greatly improved with the
launch of {\it Rossi X-ray Timing Explorer} (RXTE; Bradt, Rothschild, \&
Swank 1993). {\it RXTE} carries an all-sky monitor (ASM) that continuously 
scans the sky for transient events. Upon detection, the main instruments
can be re-programmed to observe the event within hours. To take advantage
of these unprecedented capabilities, we have established a comprehensive 
program to sample the rising phase of an outburst at a high rate, in order to 
facilitate systematic studies of the phenomenon. Luckily, we have already
caught several outbursts from different sources since the beginning of this 
year. In this Letter, we present results from observations of Aql X-1 during 
the rising phase of its recent outburst (Swank et al. 1998). 

Aql X-1 belongs to a minority group of SXTs that have been determined to 
contain neutron stars (as opposed to black holes) because they display Type~I 
X-ray bursts. It is known to experience frequent outbursts (roughly once
every year; see, e.g., Priedhorsky \& Terrell 1984; also public ASM/RXTE light 
curves). The {\it RXTE} observations of Aql X-1 during a previous outburst 
have revealed the presence of quasi-periodic oscillations (QPOs) near kHz
range in the X-ray light curves (Zhang et al. 1997), which seem to be 
common for X-ray binaries with a weakly magnetized neutron star (review by 
van der Klis 1997). The kHz QPO phenomenon is currently 
thought to be associated with the accretion disk, perhaps representing 
processes varying on dynamical time scales in the inner portion of the disk. 
There is suggestive evidence that the magnetic field might play a crucial 
role in actually producing the observed X-ray modulation (see, e.g., Zhang, Yu,
\& Zhang 1998). One of the neutron star SXTs, 4U 1608-52, is further 
classified as an ``atoll'' source, whose neutron star is thought to have 
even weaker magnetic field than a ``Z'' source (van der Klis 1995). The 
classification of Aql X-1 in this scheme is not certain.

Near the end of February 1998, the ASM detected the onset of an outburst from 
Aql X-1 (Swank et al. 1998). Fig.~1 shows a portion of the long-term ASM light
curve for the source that highlights the event. The outburst peaked at 
roughly 34 c/s (or $\sim$450 mCrab in the 1.3-12 keV band). The initial
rise lasted for about 12 days. The pending Target-of-Opportunity 
proposals of ours were triggered by the ASM alert, subsequently, a series of 
pointed {\it RXTE} observations were carried out.

\section{Observations}
The data used for this study come from a total of 16 snapshots of Aql X-1
with the {\it Proportional Counter Array} aboard {\it RXTE}, covering 
a major portion of the rising phase and the peak of the outburst. The times
when these observations were made are indicated in Fig.~1. The first 
observation took place when the ASM flux was about one quarter of the peak 
value. Each pointed {\it RXTE} observation lasted for one satellite orbit 
with an effective exposure time in the range 1.2--3.6 ks. The observations 
were conducted at a typical rate of once or twice per day, with occasional 
gaps due to scheduling constraints. Besides the two {\it Standard} 
data modes, a combination of {\it Burst Trigger} and {\it Burst Catcher} 
modes were used to zoom in onto any Type~I bursts detected; but, no X-ray 
bursts were seen. An {\it Event} mode with $\sim 122\mu s$ timing resolution 
and 64 energy bands was adopted to facilitate high-resolution timing analysis 
with a moderate energy resolution.

\section{Data Analysis and Results}
We have carried out preliminary spectral analysis, using the {\it Standard 2}
data. Throughout the entire period, the observed X-ray spectrum can be 
modeled quite well by a composite of a blackbody (BB) component (for emission 
from the neutron star surface), a multi-color disk (MCD) component (for 
emission from the accretion disk), and a thermal bremsstrahlung component
(for emission from hot plasma perhaps present in the vicinity of the neutron 
star); an emission line at $\sim$6.6 keV is also required. Fig.~2 shows the
evolution of the BB and MCD components. The spectrum remains soft, with few 
photons above $\sim$25 keV, implying the thermal origin of the emission. The 
fluxes have been computed using the best-fit parameters, with the H~I column 
density fixed at $5\times 10^{21}\mbox{ }cm^{-2}$ (Christian \& Swank 1997) 
during the fit. The discussion of physical models for such a spectrum is 
beyond the scope of this work; it will be presented in a future paper on 
detailed spectral modeling. 

For each observation, we have made light curves (with background subtracted)
from the {\it Standard 2} data (with 16-second timing resolution) in three 
energy bands: 2--5.2 keV, 5.2--8.9 keV, and 8.9--19.7 keV, from which 
two hardness ratios, 5.2--8.9 keV/2--5.2 keV (soft color) and 
8.9--19.7 keV/5.2--8.9 keV (hard color), have been calculated. The results 
are summarized in Fig.~3a, in the form of a color-color diagram for each 
observation. Fig.~3b shows the overall spectral evolution by putting together 
all the results. There are two apparent spectral transitions between three 
disjointed branches. Initially, the source is on the upper of the two left 
branches. It then seems to have made a rapid spectral transition to the lower 
one (on a time scale less than half a day; see Fig.~1). As the X-ray flux 
increases, the source evolves slightly, moving from the upper-right side to 
the lower-left side, and stays on this branch for 4 days before making another
transition to the right branch. Subsequently, the source continues its 
evolution by moving up and down the branch from left to right as the source 
flux fluctuates around the peak of the outburst.

Using the {\it Event} mode data, we have carried out FFT for every 128-second 
data segment of each observation. Individual power-density spectra (PDSs)
are then properly weighted and co-added to obtain the average PDS for 
the observation. A kHz QPO is detected only when the source is on the left 
lower branch of the color-color diagram (see Fig.~4 for an example). The QPO 
profile can be modeled well by a simple Lorentzian 
function, and the best-fit parameters are shown in Table~1. The errors are
estimated based on $\Delta \chi^2 = 1$. From the table, we can see that the 
QPO frequency varies in the range 670--870 Hz and width (FWHM) in the range 
9--20 Hz. As the flux increases, the QPO frequency also increases, and appears
to have reached a plateau at $\sim$850 Hz (following Observation 3; see 
Table~1); the fractional rms decreases. We have also performed 
FFT for different energy bands to investigate the energy dependence of the 
QPO. The results show that the fractional rms increases almost linearly with 
energy, up to $\sim$20\% in the 13.1--19.7 keV band. The QPO becomes 
unmeasurable following the second transition.

Except for the first observation, the power-density continuum is dominated by 
low-frequency ``1/f'' noises (with the power-law slope varying in the range 
-0.8-- -1.5); very little power is measured above $\sim$1 Hz (see, e.g., 
Fig.~4). 
The source shows very little variability when it is on the left upper branch
(i.e., Observation 1). It becomes more variable, following the first 
transition. The measured fractional rms (for the continuum) increases from 
$\sim$1.5\% and to $\sim$6.4\%. The variability seems to drop slightly, 
following
the second transition to the right branch, with the fractional rms staying at 
$\sim$3--4\% subsequently. 

\section{Discussion}
\subsection{Is Aql X-1 an ``Atoll'' Source?}
The phenomenology of the observed spectral evolution is very similar to that
of ``atoll'' sources (van der Klis 1995, and references therein).
In the color-color diagram (Fig.~3b), the left branches may represent so-called
``island'' states, while the curved right branch certainly shows the 
characteristics of a ``banana'' state. As the observed X-ray flux 
(presumably, the mass accretion rate) increases, the source moves from one
``island'' to another, then to the ``banana'' branch. Fig.~3a shows that 
the correlation 
between spectral states and mass accretion rate holds remarkably well in 
detail. For instance, as soon as the X-ray flux begins to drop in 
observations 15 and 16, the source reverses its motion along the branch and
moves from right to left down the ``banana''. Perhaps, Aql X-1 should be
classified as an ``atoll'' source, like its cousin 4U 1608-52. However, 
Aql X-1 also differs from 4U 1608-52 or other typical ``atoll'' sources
in that the high-frequency noise component (van der Klis 1995) is absent from 
the measured PDS for both the ``island'' states and ``banana'' state. 

For ``atoll'' sources, the physical processes that trigger the transition 
between ``island'' states themselves or between an ``island'' state and a 
``banana'' state are still not well understood. In the following, we discuss 
a physical scenario that can
account for the observed spectral and temporal evolution of Aql X-1 during
the rising phase of the outburst, based on our current knowledge of SXTs and
their outburst mechanisms.

\subsection{Transition between ``Island'' States}
It is thought that for SXTs in quiescence the mass accretion process is likely
to operate in the form of ``advection-dominated'' accretion flows (ADAFs; 
Narayan \& Yi 1994, 1995; Narayan 1997). In this regime, the compact object is
surrounded by a hot ADAF region that is sub-Keplerian and experiences a phase 
transition to the standard thin disk (Shakura \& Sunyaev 1973) thousands of 
Schwarzschild radii away (Narayan 1997). In the context of disk instability
models, when thermal instability sets in, the mass accretion rate surges,
thus, an outburst is under way. In the process, the ADAF region is cooled 
efficiently via inverse Compton scattering processes, because of the increase 
of ``seed'' photons from the disk. Consequently, the region shrinks; at the 
same time, the thin disk in the outer region fills in. Such a transition is 
likely to continue until the inner edge of the disk reaches the last 
(marginally) stable orbit (Narayan 1997). Although many details in the ADAF 
model, such as the location of the transition zone between the ADAF and the 
thin disk, still need to be worked out, the general evolution sequence has 
been shown to be followed well (see, e.g., Esin, McClintock, \& Narayan 1997). 

For SXTs that contain a neutron star, the effects of magnetic 
field must be taken into account. Unfortunately, neither do we know the
precise configuration of the magnetic field in the presence of an accretion 
disk, nor do we fully understand the disk-field interaction (see, e.g., 
Ghosh \& Lamb 1979 and Lovelace, Romanova, \& Bisnovatyi-Kogan 1995). 
Assuming a simple 
dipole field, the magnetospheric radius is given by (see, e.g., Lamb, Pethick,
\& Pines 1973 and Cui 1997)
\begin{equation}
r_m = 10^7 \xi L^{-2/7}_{x,36} M^{1/7}_{1.4} B^{4/7}_{9} R^{10/7}_6 cm
\end{equation}
where $\xi$ is a model-dependent and dimensionless number, ranging from 0.52
(Ghosh \& Lamb 1979) to $\sim$1 (Arons 1993; Ostriker \& Shu 1995; Wang 1996);
$L_{x,36}$ is the bolometric X-ray luminosity in units of $10^{36}\mbox{ }
ergs/s$; $B_9$ is the dipole field strength at the poles in units of $10^9$ G;
$M_{1.4}$ is the mass of the neutron star in units of $1.4 M_{\odot}$, and 
$R_6$ is its radius in units of 10 km.

At the beginning of an outburst, the mass accretion rate is low, so the 
magnetosphere initially extends much beyond the last stable orbit. As the 
inner edge of the disk subsequently moves in toward the neutron star, it is 
bound to encounter the magnetosphere first. We argue that it is the onset of 
this disk-field engagement that might have caused the transition between the 
two ``island'' states. Perhaps, observations 1 and 2 were made 
just before and after the engagement. For Observation 2, $L_{x,36}=5.7$ (for a 
source distance of 2.5 kpc, after being corrected for absorption), so 
$r_m = 61 \xi B_9^{4/7}\mbox{ }km$. For $r_m > 3R_s$ (the radius of 
the last stable orbit for a slowly rotating neutron star), where $R_s$ is the 
Schwarzschild radius, $B_9 > 0.062$ (or $0.20$) for $\xi=1$ (or $0.52$).

As the mass accretion rate increases, the BB temperatures increases, as shown
in Fig.~2. At the same time, the magnetosphere is pushed back further, so the 
disk extends closer to the neutron star and becomes hotter (also Fig.~2).
The intensified emission, both from the neutron star surface and accretion
disk (mostly in the 2--5.2 keV band, and also in the 5.2--8.9 keV band), 
results in the decrease of hardness ratios as the source evolves in the 
second  ``island'' state. 

\subsection{Transition between ``Island'' and ``Banana'' States}
The evolution continues until the mass accretion rate becomes so large that 
the magnetosphere is entirely inside the last stable orbit (as indicated by
the constancy of the inferred radius of the inner edge of the disk following
Observation 6; see Fig.~2). Then, the disk is once again disengaged from the 
magnetic field. This might have triggered the transition from the lower 
``island'' state to the ``banana'' state, as indicated by the disappearance
of the kHz QPO. Then, an {\it upper} limit could be derived for the magnetic 
field from equation (1). The transition is clearly under way during 
Observation 6. For this observation, $L_{x,36}=10.7$, so 
$r_m \lesssim 3R_s$ leads to $B_9 \lesssim 0.085$ (or $0.27$) for $\xi=1$ (or 
$0.52$). 

In the ``banana'' state, the observed spectral evolution implies that the soft
fluxes (both from the disk and the neutron star surface) shift more toward the
the middle energy band (5.2-8.9 keV), as the mass accretion rate increases, 
and the hard flux increases even more. 

\subsection{Summary}
We have proposed a simple explanation for the observed spectral and temporal 
evolution of Aql X-1 during the rising phase of the outburst. It remains to
be seen whether or not the model also applies to ``atoll'' sources in general.
Progress can be made by carefully studying ``atoll'' sources in a similar way 
(i.e., correlating the occurrence of kHz QPOs with distinct spectral states). 
The model is almost certainly over-simplified, but seems to be qualitatively
valid.

It has been speculated that kHz QPOs might originate in the inhomogeneity (or 
``hot spots'') at the inner edge of the disk and the QPO frequency might 
simply be the local Keplerian frequency or the beat frequency between the 
Keplerian motion and the spin of the neutron star (see van der Klis 1997 for 
a review). If so, the occurrence and 
disappearance of the kHz QPO for Aql X-1 would strongly suggest that the 
magnetic field plays an essential role in producing such ``hot spots'' by 
interacting with the disk. With limited statistics (Table~1), the kHz QPO 
frequency does seem to level off as the mass accretion rate continues to 
increase, perhaps indicating that the inner edge of the disk has reached
(or approached close to) the last stable orbit as early as Observation 3. 
This is not inconsistent with the spectral results (see Fig.~2).

It is instructive to compare this outburst with previous ones. A spectral 
transition was also observed near the end of the decaying phase of the first
1997 outburst (Zhang, Yu, \& Zhang 1998). It appears to be similar to the 
first transition observed here, although it started at a lower flux level. 
Interestingly enough, the kHz QPO was also detected at lower fluxes 
(Zhang et al. 1998). Although it is true that the peak intensity of the 
current outburst is higher than the previous one, it is not clear how this 
would affect the fundamental properties of the system. It is possible, 
however, that the disk structure might be different between the rising and 
decaying phases, which could result in hysteresis. More studies are clearly 
required to address this issue.  

\acknowledgments
This work is supported partially by NASA through Contract NAS5-30612. We 
would like to thank Alan Harmon and Jean Francois Olive for helpful 
discussions.

\clearpage

\begin{deluxetable}{lccc}
\tablecolumns{5}
\tablewidth{0pc}
\tablecaption{Observed kHz QPO in Aql X-1}
\tablehead{
\colhead{Observation}&\colhead{Frequency (Hz)}&\colhead{FWHM (Hz)} &\colhead{Fractional RMS (\%)}}
\startdata
2 & $677.0^{+0.4}_{-0.3}$ & $9.1^{+0.7}_{-0.9}$ & $6.3^{+0.2}_{-0.2}$ \\
3 & $856.3^{+1.1}_{-1.0}$ & $19.7^{+2.5}_{-2.0}$ & $5.7^{+0.3}_{-0.2}$ \\
4 & $839.5^{+0.5}_{-0.4}$ & $11.0^{+1.1}_{-0.9}$ & $5.7^{+0.2}_{-0.2}$ \\
5 & $871.4^{+1.0}_{-1.0}$ & $11.8^{+3.0}_{-2.4}$ & $3.3^{+0.3}_{-0.3}$ \\
\enddata
\end{deluxetable}

\clearpage
\begin{figure}
\psfig{figure=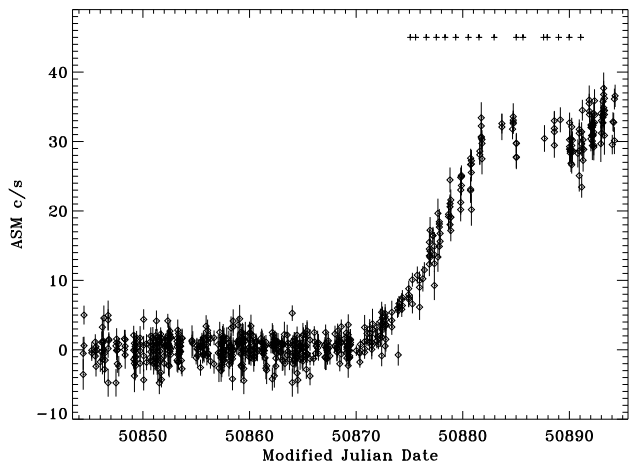,height=9cm,angle=90}
\caption{ASM light curve for Aql X-1. Each data point represents a 90-second
measurement. Note that the outburst started roughly at MJD 50870 (2/26/98).
As a reference, the Crab Nebula produces about 75 c/s. The crosses indicate 
the times when the pointed {\it RXTE} observations were carried out. The 
pointed observations are numbered sequentially, starting from Observation 1,
throughout the text. }
\end{figure}
\begin{figure}
\psfig{figure=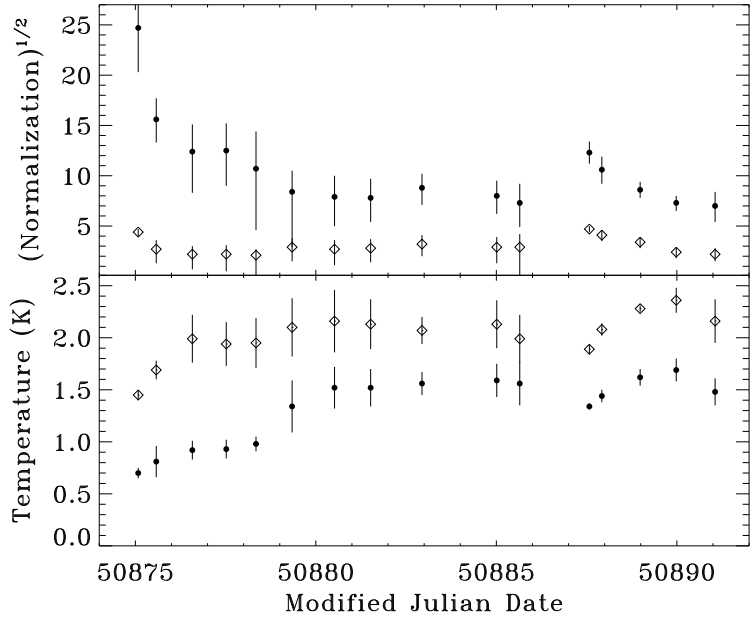,height=9cm}
\caption{Evolution of the BB (diamonds) and MCD (bullets) components. For MCD,
the temperature shown is for the inner edge of the accretion disk. Note that
the upper panel essentially shows $R_{BB}/D$ and $R_{in} (\cos i)^{1/2}/D$ for 
BB and MCD, respectively, where $R_{BB}$ is the radius of the BB emission 
region in units of km; $R_{in}$ is the radial distance 
to the inner edge of the disk in units of km; $i$ is the inclination angle of 
the disk; and $D$ is the distance to the source in units of 10 kpc. }
\end{figure}
\begin{figure}
\psfig{figure=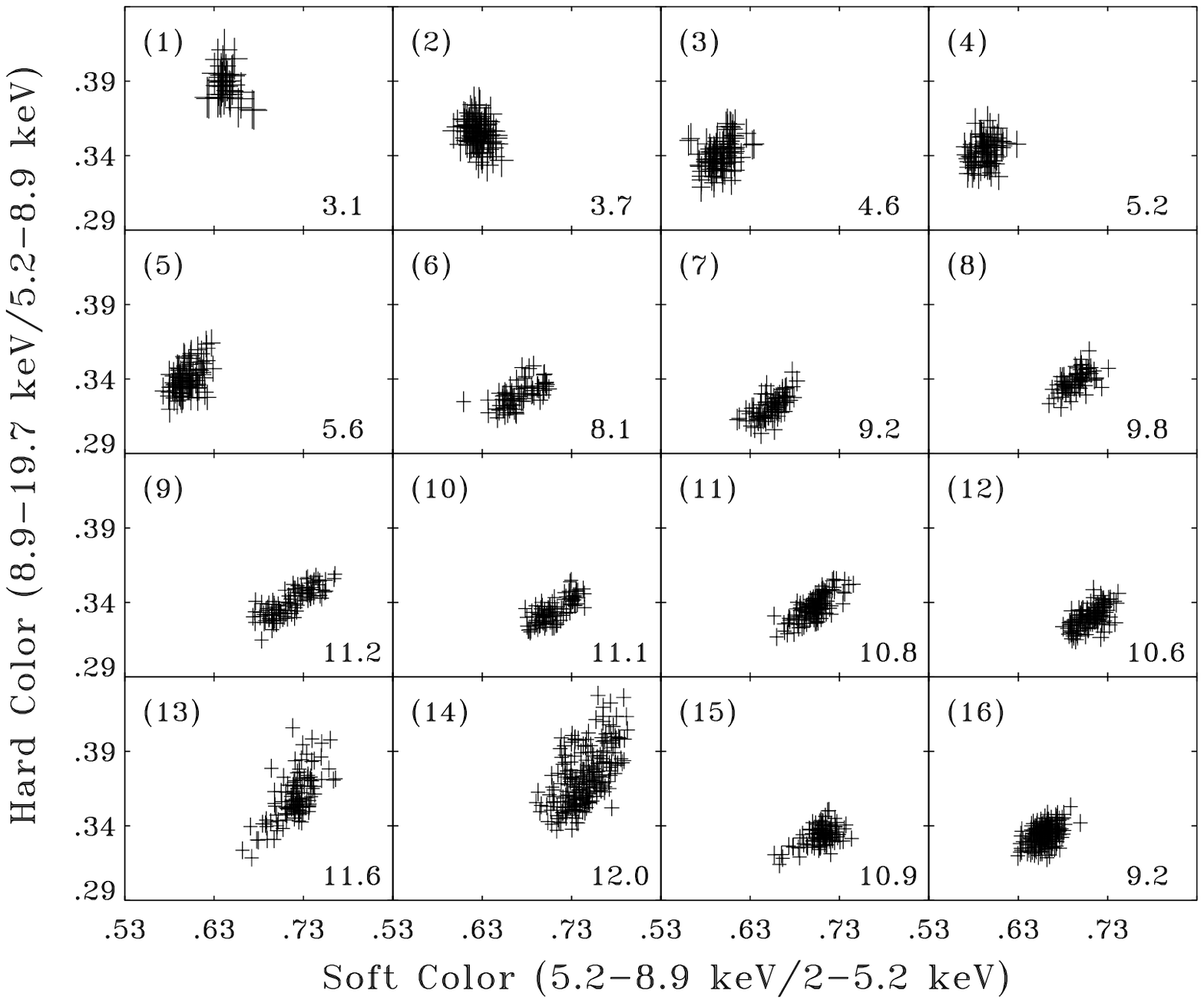,height=9cm}
\psfig{figure=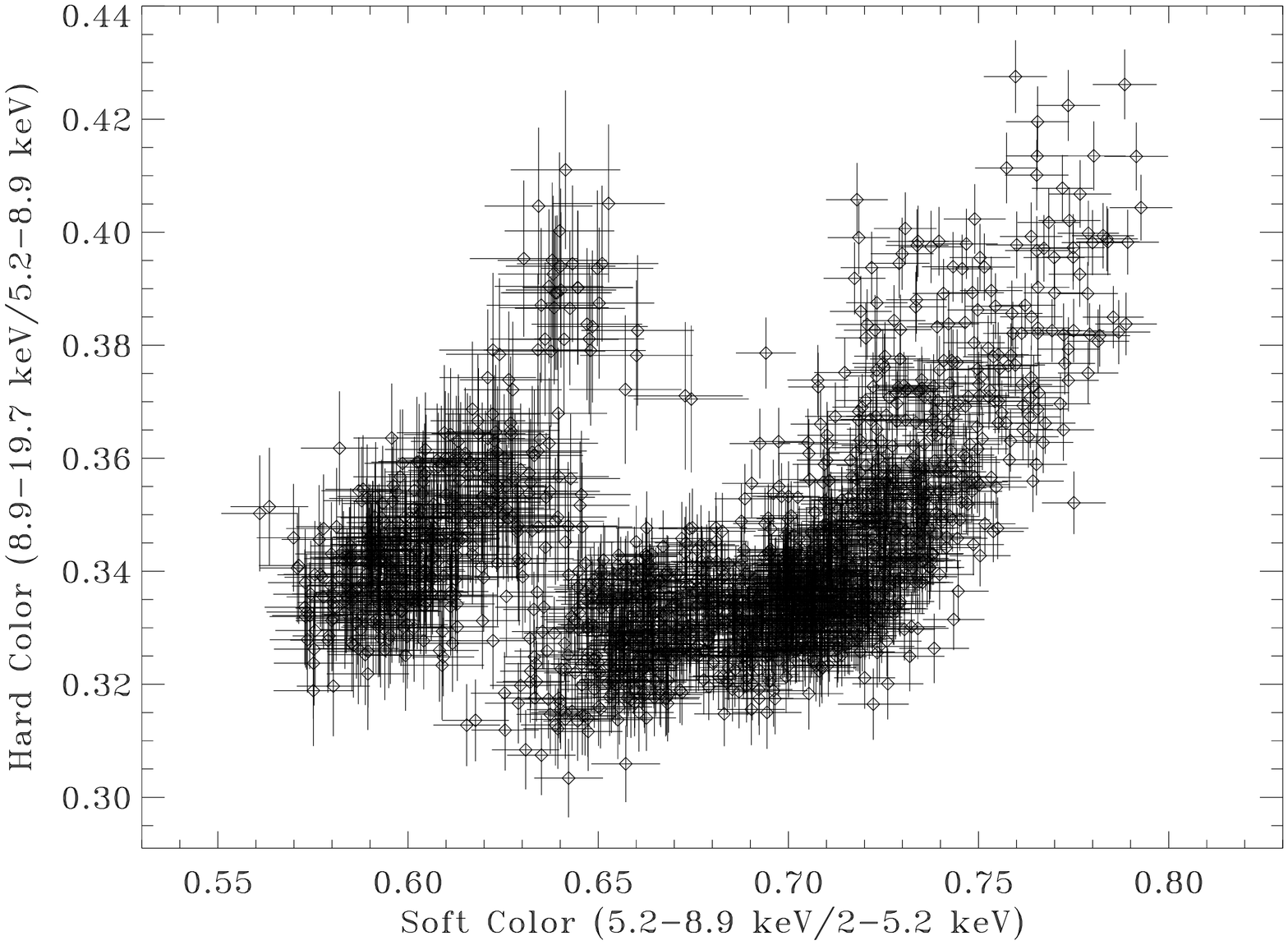,height=9cm,angle=90}
\caption{(a) Color-color diagram for each observation. Each data point 
represents a 16-second measurement. The measured 2-25 keV fluxes in units 
of $10^{-9}\mbox{ }ergs\mbox{ }cm^{-2}\mbox{ }s^{-1}$ are also shown at the 
lower right-hand corner of each panel. Note that the source moves up and 
down the right branch of the overall color-color diagram (bottom panel) as 
the X-ray flux fluctuates around the peak of the outburst. (b) Overall 
color-color diagram. Note the transitions between three distinct 
branches. The kHz QPO is only detected when the source is on the left lower
branch. }
\end{figure}
\begin{figure}
\psfig{figure=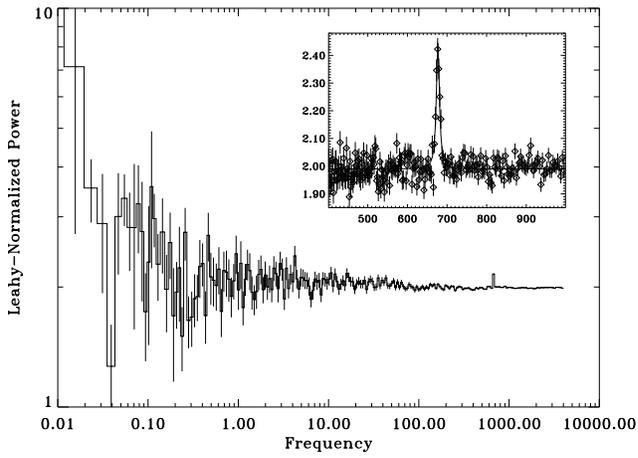,height=9cm,angle=90}
\caption{A sample power-density spectrum (taken from Observation 2). Note 
the presence of very low-frequence noises and a kHz QPO. The inset shows the
QPO profile in detail, with the solid line representing the best-fit 
Lorentzian function. }
\end{figure}

\end{document}